# Design of a Rectangular Linear Microstrip Patch Antenna Array for 5G Communication

Muhammad Asfar Saeed
*Faculty of engineering*
*University of Greenwich*
Medway, United Kingdom
M.A.Saeed@greenwich.ac.uk

Augustine O. Nwajana
*Faculty of engineering*
*University of Greenwich*
Medway, United Kingdom
a.o.nwajana@greenwich.ac.uk

*Abstract*— This paper presents the design and characterization of a rectangular microstrip patch antenna array optimized for operation within the Ku-band frequency range. The antenna array is impedance-matched to 50Ω and utilizes a microstrip line feeding mechanism for excitation. The design maintains compact dimensions, with the overall antenna occupying an area of 29.5x7 mm. The antenna structure is modelled on an R03003 substrate material, featuring a dielectric constant of 3, a low-loss tangent of 0.0009, and a thickness of 1.574 mm. The substrate is backed by a conducting ground plane, and the array consists of six radiating patch elements positioned on top. Evaluation of the designed antenna array reveals a resonant frequency of 18GHz, with a -10 dB impedance bandwidth extending over 700MHz. The antenna demonstrates a high gain of 7.51dBi, making it well-suited for applications in 5G and future communication systems. Its compact form factor, cost-effectiveness, and broad impedance and radiation coverage further underscore its potential in these domains.

*Keywords—RO3003, six elements, Radiofrequency, Ku-band, radiations,*

I. INTRODUCTION

In the relentless search for ever faster and more reliable communication systems, the utilization of higher frequency bands, particularly in the millimetre-wave spectrum, has become increasingly imperative. Among these bands, the 18GHz frequency range stands out as a promising candidate for advanced communication applications, particularly in the context of 5G and beyond [1]. This research paper aims to search into the design, analysis, and performance evaluation of a linear array operating at 18GHz. The array configuration, with its ability to dynamically control individual elements, offers significant potential for enhancing communication links through beamforming and spatial diversity techniques. The motivation behind this investigation stems from the urging need for communication systems that can accommodate the increasing demands of the modern community. With the advent of technologies such as ultra-high-definition video streaming, and the Internet of Things (IoT), there exists a critical need for communication set-ups capable of supporting extraordinary data rates, ultra-low latency, and massive device connectivity.

The 18GHz frequency band has been considered as a promising frequency range to meet these demands. Its relatively high frequency allows large bandwidth allocations, enabling the transmission of extensive amounts of data at high speed. Moreover, the shorter wavelengths associated with this frequency band enable the design of compact antenna arrays [2], making it well-suited for integration in compact electronic devices. In literature, researchers have explored many techniques to achieve beamforming [4] [5] [6]. The most widely used technique is to expand the radiating patch width, it increases the current distribution and enables the wave propagation to increase [7]. the shared coupling between radiating elements within an array is responsible for achieving wide-angle beamforming in both linear and phased arrays [8]. Some of the techniques involved include adding multilayers of dielectric substrate [9]. A new circular-polarization waveguide antenna array utilizing innovative beamforming techniques is introduced. The design employs an open quad-ridged waveguide fed by straight slots cut into the broad wall of a rectangular waveguide to achieve circular polarization [10]. A high-gain antenna is capable of directional pointing is developed by integrating beamforming capabilities into a single microstrip array antenna operating at millimetre-wave frequencies as presented in [11]. The antenna structure consists of a substrate made of alumina thin film, upon which a microstrip antenna array of defined dimensions and configuration is etched. Other techniques involve tapered antenna arrays [12]. While the methods described above enable beamforming, they exhibit certain limitations such as complex antenna structures, challenges in beam scanning accuracy, inefficiency, and fluctuations in antenna gain. These issues need to be addressed for improved performance and reliability in practical applications.

Following the introduction, this paper is organized into three main sections. Section II details the design and development process of the proposed microstrip linear array antenna, focusing on its construction and key features. Section III presents an in-depth analysis of the antenna's performance, such as gain, efficiency, and beamforming capabilities. Finally, Section IV summarizes the findings and draws conclusions based on the outcomes of the study. Each section contributes to a comprehensive understanding of the antenna's design, functionality, and practical implications in modern communication systems.

## II. DESIGN AND DEVELOPMENT OF MPA ARRAY

A microstrip patch antenna array has been developed, featuring six radiating patches arranged in a linear configuration atop a rectangular substrate. The ground layer covers the bottom of the substrate, ensuring proper antenna operation. The antenna structure occupies a total area of 29.5x7 mm and utilizes thin transmission lines to interconnect the radiating patches, facilitating the transmission of current between them. The antenna is excited via an edge-mounted connector, with a microstrip line serving as the means of excitation for the linear array. This excitation results in the generation of a fan beam pattern for radiation, characterized by constructive and destructive interference effects arising from the linear arrangement of the radiating patches. These interference phenomena contribute to the shaping of the antenna's radiation pattern, allowing for directional beamforming capabilities. Ro3003 is utilized as a dielectric material as it offers a Low Dielectric Loss RO3003 exhibits low dielectric loss, ensuring efficient signal transmission and minimal energy dissipation. Moreover, this material has a High Dielectric Constant, RO3003 enables compact antenna designs by reducing wavelength, leading towards the contribution of smaller antenna dimensions which is the objective of the design. It maintains stable electrical properties over a wide range of frequencies and temperatures, ensuring consistent antenna performance in different applications where the electronic device is exposed to an open environment in high temperatures [13]. In addition, the substrate's mechanical stability enhances the antenna's durability and reliability, crucial for long-term operation and it offers easy integration to the circuit [12]. The designed microstrip patch antenna array is fed using a microstrip line feeding method it can be easily integrated into the antenna structure, simplifying the overall design process, it also enables precise control over the phase of the signals fed to each antenna element, facilitating beamforming and steering capabilities [14]. Microstrip lines exhibit low transmission losses, ensuring efficient signal delivery and maximizing antenna performance. It also contributes to the compactness of the microstrip patch antenna array as it can be fabricated on the same substrate. The dimensions of the designed antenna are illustrated in Table 1. The proposed antenna design is modelled and simulated using CST Microwave Studio®, a software platform that employs the Finite Integration Technique (FIT) to solve Maxwell's equations [16]. This simulation approach allows for accurate analysis of electromagnetic behaviour, including wave propagation, reflection, and diffraction, within the antenna structure. CST Microwave Studio® using the FIT method provides high-fidelity simulations, capturing complex electromagnetic interactions with precision. The FIT solver enables full-wave analysis of the antenna's performance, accounting for all electromagnetic effects including near-field interactions and far-field radiation. The simulation tool supports a wide frequency range, accommodating the millimetre-wave frequency of the proposed antenna design at 18GHz.

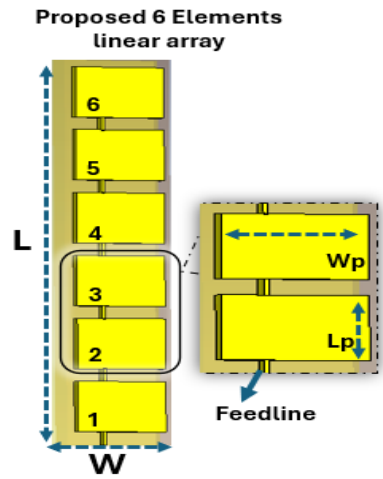

Figure 1 Proposed MPA array.

## III. MPA ARRAY PERFORMANCE AND EVALUATION

The designed linear MPA array is evaluated based on the free space important parameters like reflection coefficient, impedance matching, current distribution, radiation pattern, and Gain.

### A. RELECTION COEFFICEINT

S11 measures the amount of power reflected by the antenna relative to the power incident upon it. A low S11 value indicates good impedance matching between the antenna and the transmission line, maximizing power transfer efficiency. The designed antenna has achieved a -10db impedance bandwidth of 1GHz at the resonant frequency and return loss lesser than -16 dB indicates minimal reflection within the transmission line which is noticeable in the figure 2.

### B. IMPEDANCE MATCHING

The designed antenna is fed using the microstrip line feeding method and the results shown in the figure 5 illustrate that the designed antenna is perfectly matched at 50Ω. Impedance matching ensures that the antenna's input impedance matches the characteristic impedance of the transmission line, minimizing signal reflections and maximizing power transfer it is obvious from the figure that maximum current is coming from the transmission line.

### C. Current distribution

It is evident from the simulated results that a progressive wave pattern is being formed. Analysing the distribution of current from figure 3 across the antenna elements provides insights into how electromagnetic energy is distributed and radiated. This helped in understanding the antenna's radiation characteristics and optimizing its design for desired performance.

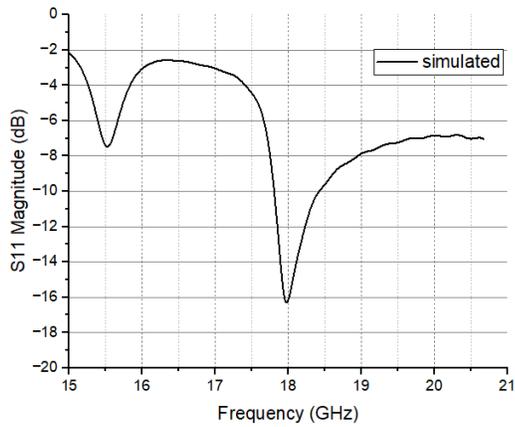

Figure 2 Simulated Reflection coefficient response.

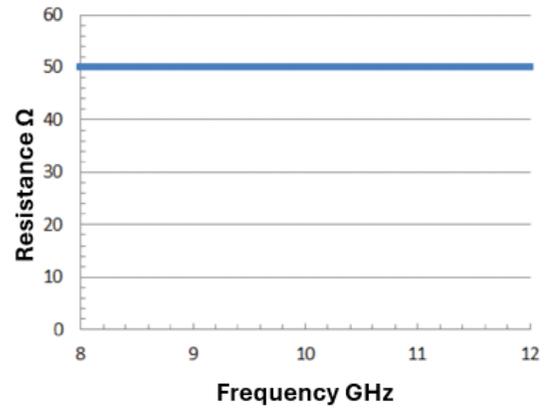

Figure 5 Reference Impedance

*D. Radiation Pattern*

The 2D radiation pattern of the designed antenna array is illustrated in the figure 4. It is evident from the figure that the designed antenna obtained a covered in the form of a beam, it also shows that the designed linear array has good coverage in both the E-plane and H-plane. The radiation pattern describes the directional distribution of radiated electromagnetic energy from the antenna. It includes parameters such as beamwidth, directivity, and sidelobe levels, which determine the antenna's coverage area and spatial focusing capabilities which are noticeable in the simulated results.

*E. GAIN*

The antenna gain is evident from the simulated results and the designed antenna has achieved a higher gain of 7.511dB which quantifies the ratio of radiation intensity in a particular direction to that of an isotropic radiator (which radiates uniformly in all directions). Higher gain indicates greater radiation concentration in the desired direction, improving communication range and signal strength.

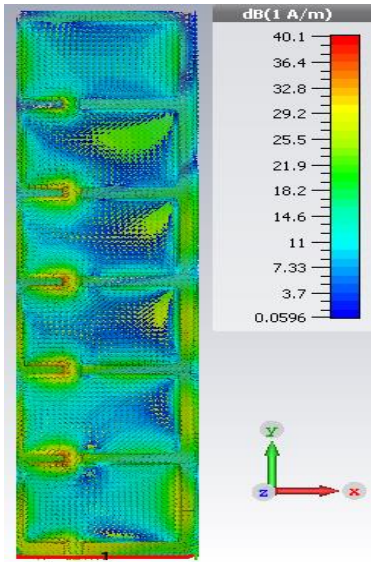

Figure 3 Current distribution of the proposed MPA array.

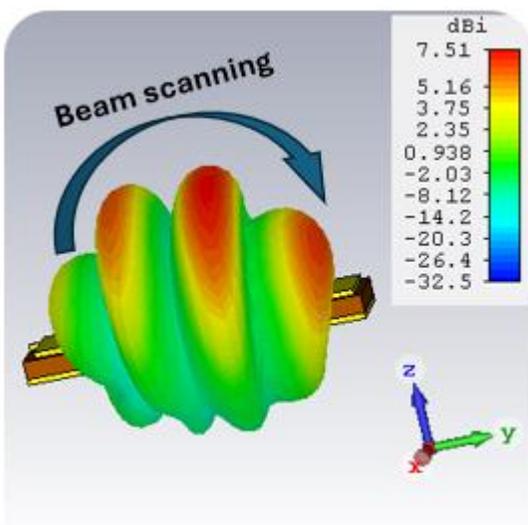

Figure 4 3-D Far-field radiation pattern

IV. CONCLUSION

In conclusion, the design and simulation of a microstrip patch antenna (MPA) linear array operating at 18GHz frequency, utilizing RO3003 as the substrate material, has yielded promising results. The antenna, containing six radiating patches arranged in a rectangular configuration, exhibits satisfactory performance across key parameters. Through simulation, important parameters such as reference impedance, reflection coefficient, current distribution, and gain have been thoroughly analysed and found to meet design expectations. Notably, the designed MPA array demonstrates the ability to form a beamforming pattern, contributing to enhanced gain and efficiency. This beamforming capability allows for the directional concentration of electromagnetic energy, improving the antenna's effectiveness in communication applications.

Overall, the successful modelling and simulation of the MPA array validate its potential for practical implementation in various communication systems operating at 18GHz. This research lays the foundation for continued exploration and development of high-frequency antenna technologies, contributing to advancements in wireless communication and beyond.

TABLE I. DIMENSIONS OF THE MPA ARRAY

| ANTENNA ELEMENTS | PARAMETERS | DIMENSIONS (MM) |
|---|---|---|
| PATCH | Length (L) | 3.85 |
|  | Width (W) | 5.89 |
| GROUND | Length (GL) | 29.50 |
|  | Width (GW) | 7 |
|  | Thickness (t) | 0.5 |
| FEEDLINE | Length (FL) | 1 |
|  | Width (FW) | 0.20 |
| SUBSTRATE | Length (L) | 29.50 |
|  | Width (W) | 7 |
|  | Thickness (h) | 1.574 |
|  | Permittivity ($\varepsilon$) | 3 |

TABLE II COMPARISON OF THE MPA ARRAY

| REF | ANTENNA TYPE | RESONANT FREQUENCY | RETURN LOSS | GAIN |
|---|---|---|---|---|
| [12] | PLANAR ARRAY | 4-7GHz | -22 | 6.5 dBi |
| [13] | COUPLED ARRAY | 28.5GHz | -14 | 1.51 dBi |
| [8] | PHASED ARRAY | 25-33GHz | -23 | 5 dBi |
| PROPOSED | LINEAR ARRAY | 18GHz | -16 | 7.91 dBi |

ACKNOWLEDGMENT

The paper is written under the supervision of Augustine O. Nwajana during the PhD project at research RF and antenna research lab, funded by the University of Greenwich.